\documentclass[final]{IEEEtran}

\usepackage{graphicx}
\usepackage{subfigure}

\usepackage{amsmath}
\usepackage{amssymb}
\usepackage{algorithmic}
\usepackage{algorithm}

\newtheorem{theorem}{Theorem}
\newtheorem{lemma}{Lemma}

\newtheorem{remark}{Remark}

\newcommand{\dotcup}{\ensuremath{\mathaccent\cdot\cup}}

\begin{document}

\title{On High Spatial Reuse Link Scheduling in STDMA Wireless Ad Hoc Networks}

\author{Ashutosh~Deepak~Gore\thanks{A.D. Gore and A. Karandikar are
    with the Information Networks Laboratory, Department of Electrical
    Engineering, Indian Institute of Technology - Bombay, Mumbai
    400076, India.  Email:
    \{adgore,karandi\}@ee.iitb.ac.in},~Srikanth~Jagabathula\thanks{S.
    Jagabathula is with the Laboratory for Information and Decision
    Systems, Massachusetts Institute of Technology, Cambridge, MA
    02139, USA.  Email: jskanth@mit.edu}~and~Abhay~Karandikar }

\maketitle

\begin{abstract}
Graph-based algorithms for point-to-point link scheduling in Spatial
reuse Time Division Multiple Access (STDMA) wireless ad hoc networks
often result in a significant number of transmissions having low
Signal to Interference and Noise density Ratio (SINR) at intended
receivers, leading to low throughput. To overcome this problem, we
propose a new algorithm for STDMA link scheduling based on a graph
model of the network as well as SINR computations.  The performance of
our algorithm is evaluated in terms of spatial reuse and computational
complexity.  Simulation results demonstrate that our algorithm
achieves better performance than existing algorithms.
\end{abstract}


\section{Introduction}

\PARstart{A}{} wireless ad hoc network consists of a finite number of
radio units (nodes) that are geographically distributed in a terrain
without any preplanned or fixed infrastructure. They communicate with
each other via the untethered and broadcast wireless medium.  In order
to use the scarce and expensive wireless spectrum efficiently, we need
to exploit channel spatial reuse, i.e., allow concurrent communication
between source-destination pairs which are ``reasonably" far from each
other using either the same time slot or frequency band.

A commonly used scheme for channel reuse is Spatial Time Division
Multiple Access (STDMA), in which time is divided into fixed-length
slots that are organized cyclically. An STDMA schedule describes the
transmission rights for each time slot in such a way that
communicating pairs assigned to the same slot do not collide.  STDMA
scheduling algorithms can be categorized into link scheduling
 and broadcast/node scheduling algorithms
\cite{ramanathan_lloyd__scheduling_algorithms}. In a wireless ad hoc
network, a link is an ordered pair of nodes $(t,r)$, where $t$ is a
transmitter and $r$ is a receiver. In link scheduling, the
transmission right in every slot is assigned to certain links. On the
other hand, in broadcast scheduling, the transmission right in every
slot is assigned to certain nodes. Thus, there is no apriori binding
of transmitter and receiver and the packet transmitted can be received
by every neighbor.  Link scheduling is suitable for unicast traffic,
while broadcast scheduling is suitable for broadcast traffic. In this
paper, we will concentrate on link scheduling for STDMA
networks. Specifically, we consider centralized link scheduling, i.e.,
the link transmission schedule is computed by a central
entity. Centralized scheduling is applicable for scenarios where the
time scale of topology change is much larger than the duration of the
schedule.

\subsection{Related Work}

The concept of STDMA for multihop wireless ad hoc networks was
formalized in \cite{nelson_kleinrock__spatial_tdma}. Centralized
algorithms \cite{hajek_sasaki__link_scheduling}
\cite{funabiki_takefuji__parallel_algorithm} as well as distributed
algorithms \cite{chlamtac_lerner__fair_algorithms}
\cite{chlamtac_pinter__distributed_nodes} have been proposed for
generating reuse schedules. The problem of determining an optimal
minimum-length STDMA schedule for a general multihop ad hoc network is
NP-complete for both link and broadcast scheduling
\cite{ramanathan_lloyd__scheduling_algorithms}.  In fact, this is
closely related to the problem of determining the minimum number of
colors to color all the edges (or vertices) of a graph under certain
adjacency constraints.  However, most wireless ad hoc networks can be
modeled by planar or close-to-planar graphs and thus near-optimal edge
coloring algorithms can be developed for these restricted classes of
graphs.

A significant work in STDMA link scheduling is reported in
\cite{ramanathan_lloyd__scheduling_algorithms}, in which the authors
show that tree networks can be scheduled optimally, oriented graphs
can be scheduled near-optimally and arbitrary networks can be
scheduled such that the schedule is bounded by a length proportional
to the graph thickness\footnote{The thickness of a graph is the
minimum number of planar graphs into which the given graph can be
partitioned.} times the optimum number of colors.

A probabilistic analysis of the throughput performance of graph-based
scheduling algorithms under the physical interference model is derived
in \cite{behzad_rubin__performance_graph}. The authors determine the
optimal number of simultaneous transmissions by maximizing a lower
bound on the physical throughput and subsequently propose a truncated
graph-based scheduling algorithm that provides probabilistic
guarantees for network throughput.

In \cite{guo_roy__spatial_reuse}, the authors present an analytical
framework to investigate co-channel spatial reuse in dense wireless ad
hoc networks based on path loss and log-normal shadowing models for
a 1-D infinite regular chain topology and a 2-D infinite
hexagonally-tessellated topology. They derive the minimum ratio of
inter-transmitter distance to transmitter-receiver distance, while
still maintaining desirable Signal to Interference and Noise
density Ratio (SINR) at the receivers. Their results
demonstrate that increasing transmission power improves spatial reuse
in ambient noise dominated environments. However, in co-channel
interference limited scenarios, increasing transmission power has
little effect on spatial reuse.

The performance of centralized graph-based and interference-based
STDMA scheduling via simulations is evaluated and compared in
\cite{gronkvist_hansson__comparison_between}.  To generate a
graph-based conflict-free schedule, the authors use a two-level graph
model with certain SINR threshold values chosen based on heuristics
and examples. To generate an interference-based conflict-free
schedule, the authors employ a method suggested in
\cite{somarriba__multihop_packet} which describes heuristics based on
two path loss models, namely terrain-data based ground wave propagation
model and Vogler's five knife-edge model.

In \cite{brar_blough_santi__computationally_efficient}, the authors
investigate throughput improvement in an 802.11-like wireless mesh
network with Carrier Sense Multiple Access with Collision Avoidance
(CSMA/CA) channel access scheme replaced by STDMA. For a successful
packet transmission, they mandate that two-way communication be
successful. Under this `extended physical interference model', they
present a greedy algorithm which computes the transmission schedule in
a centralized manner. Assuming a uniform random node distribution and
using results from occupancy theory, they derive an approximation
factor for the length of this schedule relative to the shortest
schedule.

Link scheduling for power-controlled STDMA
networks under the physical interference model is analyzed in
\cite{moscibroda_wattenhofer__complexity_connectivity}. The authors
define the scheduling complexity as the minimum number of time slots
required for strong connectivity of the network. They develop an
algorithm employing non-linear power assignment and show that its
scheduling complexity is polylogarithmic in the number of nodes.

In a related work
\cite{moscibroda_wattenhofer_zollinger__topology_control}, the authors
investigate the time complexity of scheduling a set of communication
requests in an arbitrary network.  They consider a `generalized
physical model' wherein the actual received power of a signal can
deviate from the theoretically received power by a
multiplicative factor. Their algorithm successfully schedules all
links in time proportional to the squared logarithm of the number of
nodes times the static interference measure
\cite{rickenbach_schmid__robust_interference}.

In \cite{kim_lim_hou__improving_spatial}, the authors investigate the
tradeoff between the average number of concurrent transmissions
(spatial reuse) and sustained data rate per node for an 802.11
wireless network.  Assuming that the channel data rate is given by the
Shannon capacity, they show that spatial reuse depends only on the
ratio of transmit power to carrier sense threshold. Keeping the
carrier sense threshold fixed, they propose a distributed power and
rate control algorithm based on interference measurement and evaluate
its performance via simulations.

In \cite{lim_lim_hou__coordinate_based}, the authors investigate
mitigation of inter-flow interference in an 802.11e wireless mesh
network from a temporal-spatial diversity perspective.  Measurements
of received signal strengths are used to construct a virtual
coordinate system to identify concurrent transmissions with minimum
inter-flow interference. Based on this new coordinate system, the
gateway node determines the scheduling order for downlink frames of
different connections. Through extensive simulation with real-life
measurement traces, the authors demonstrate throughput improvement
with their algorithms.

In \cite{kodialam_nandagopal__characterizing_achievable}, the authors
consider wireless mesh networks with half duplex and full duplex
orthogonal channels, wherein each node can transmit to at most one
node and/or receive from at most $k$ nodes ($k \geqslant 1$) during
any time slot. They investigate the joint problem of routing flows and
scheduling link transmissions to analyze the achievability of a given
rate vector between multiple source-destination pairs. The scheduling
problem is solved as an edge-coloring problem on a multi-graph and the
necessary conditions from scheduling problem lead to constraints on
the routing problem, which is then formulated as a linear optimization
problem. Correspondingly, the authors present a greedy coloring
algorithm to obtain a 2-approximate solution to the chromatic index
problem and describe a polynomial time approximation algorithm to
obtain an $\epsilon$-optimal solution of the routing problem using the
primal dual approach. Finally, they evaluate the performance of their
algorithms via simulations.

\subsection{Contributions of our Work}

In most STDMA algorithms, a link schedule is usually determined from a
graph model of the network
\cite{ramanathan_lloyd__scheduling_algorithms}
\cite{behzad_rubin__performance_graph}
\cite{gronkvist_hansson__comparison_between}. However, graph-based
scheduling algorithms assume a limited knowledge of the interference
and result in low network throughput. On the other hand, SINR-based
scheduling algorithms
\cite{brar_blough_santi__computationally_efficient}
\cite{moscibroda_wattenhofer__complexity_connectivity}
\cite{moscibroda_wattenhofer_zollinger__topology_control}
\cite{kim_lim_hou__improving_spatial} require a complete knowledge of
the interference and lead to higher throughput.  Existing literature
on SINR-based STDMA link scheduling consider system models which are
different from our system model. For example,
\cite{brar_blough_santi__computationally_efficient}
\cite{kim_lim_hou__improving_spatial} consider a variant of 802.11
wireless networks, \cite{lim_lim_hou__coordinate_based}
\cite{kodialam_nandagopal__characterizing_achievable} consider
wireless mesh networks and
\cite{moscibroda_wattenhofer__complexity_connectivity}
\cite{moscibroda_wattenhofer_zollinger__topology_control} assume
non-uniform transmit power at all nodes.

In this paper, we consider an STDMA wireless ad hoc network with
uniform transmit power at all nodes and propose a link scheduling
algorithm based on the graph model as well as SINR computations. We
introduce spatial reuse as an important performance metric and argue
that a high value of spatial reuse directly translates to high
long-term network throughput. We show that the proposed algorithm has
low computational complexity and high spatial reuse compared to
existing algorithms.

The rest of the paper is organized as follows. In Section \ref{xkybv},
we describe our system model along with the physical and protocol
interference models, discuss the limitations of graph-based scheduling
algorithms, formulate the problem and summarize the differences
between our work and existing work in SINR-based scheduling
algorithms. Section \ref{lxvym} describes the proposed link scheduling
algorithm. The performance of our algorithm is evaluated in Section
\ref{cymoe} and its computational complexity is derived in Section
\ref{mbxgt}. We conclude and suggest directions for future work in
Section \ref{eiovv}.

\section{System Model}
\label{xkybv}

Consider an STDMA wireless ad hoc network with $N$ static nodes
(wireless routers) in a two-dimensional plane.  During a time slot, a
node can either transmit, receive or remain idle. We assume
homogeneous and backlogged nodes. Let:

\begin{eqnarray*}
(x_j,y_j) &=& 
\mbox{Cartesian coordinates of $j^{th}$ node} \;=:\;{\mathbf r}_j\\
P &=& \mbox{transmission power of every node}\\
N_0 &=& \mbox{thermal noise density}\\
D(j,k) &=& \mbox{Euclidean distance between nodes $j$ and $k$}
\end{eqnarray*}

We do not consider fading and shadowing effects. The received signal
power at a distance $D$ from the transmitter is given by
$\frac{P}{D^{\alpha}}$, where $\alpha$ is the path loss factor.

A link schedule effectively assigns sets of links to time slots.
Specifically, a link schedule for the STDMA network is denoted by
$\Psi(C,{\mathcal S}_1,\cdots,{\mathcal S}_C)$, where
\begin{eqnarray*}
C &=& \mbox{number of slots in the link schedule}\\
{\mathcal S}_i 
 &=& \mbox{set of transmitter-receiver pairs which can}\\
&& \mbox{communicate concurrently in the $i^{th}$ slot}\\
 &:=& \{t_{i,1}\rightarrow r_{i,1},\cdots,t_{i,M_i}\rightarrow r_{i,M_i}\}
\end{eqnarray*}
where $t_{i,j}\rightarrow r_{i,j}$ denotes a packet transmission from
node $t_{i,j}$ to node $r_{i,j}$ in the $i^{th}$ slot\footnote{A node
is generically denoted by $j$, $j=1,\ldots,N$. However, we have used
the notation $t_{i,j}$ to denote a node {\it transmitting} in the
$i^{th}$ slot. Similarly, a node {\it receiving} in the $i^{th}$ slot
is denoted by $r_{i,j}$.}. Note that $t_{i,j}, r_{i,j} \in
\{1,\ldots,N\}$ and $M_i = |{\mathcal S}_i|$.  The SINR at receiver
$r_{i,j}$ is given by
\begin{eqnarray}
{\mbox{SINR}}_{r_{i,j}} &=& 
 \frac{\frac{P}{D^{\alpha}(t_{i,j},r_{i,j})}}{N_0+\sum_{\stackrel{k=1}{k\neq j}}^{M_i}\frac{P}{D^{\alpha}(t_{i,k},r_{i,j})}}
\label{xvmyw}
\end{eqnarray}
We define the signal to noise ratio (SNR) at receiver $r_{i,j}$ by
\begin{eqnarray}
{\mbox{SNR}}_{r_{i,j}} &=& \frac{P}{N_0D^{\alpha}(t_{i,j},r_{i,j})}
\label{xklfy}
\end{eqnarray}

\subsection{Physical and Protocol Interference Models}

According to the {\it physical interference model}
\cite{gupta_kumar__capacity_wireless}, $t_{i,j}\rightarrow r_{i,j}$ is
successful if and only if (iff) the SINR at receiver $r_{i,j}$ is
greater than or equal to a certain threshold $\gamma_c$, termed as the
communication threshold.
\begin{eqnarray}
\frac{\frac{P}{D^\alpha(t_{i,j},r_{i,j})}}{N_0+\sum_{\stackrel{k=1}{k\neq j}}^{M_i} \frac{P}{D^\alpha(t_{i,k},r_{i,j})}} \geqslant \gamma_c
\label{mnmnx}
\end{eqnarray}

\noindent According to the {\it protocol interference model}
\cite{gupta_kumar__capacity_wireless},
$t_{i,j}\rightarrow r_{i,j}$ is successful if:
\begin{enumerate}
\item
the SNR at receiver $r_{i,j}$ is no less than the communication
threshold $\gamma_c$.  From (\ref{xklfy}), this translates to
\begin{eqnarray}
D(t_{i,j},r_{i,j}) &\leqslant& \left(\frac{P}{N_0 \gamma_c}\right)^\frac{1}{\alpha}
 \;\;=:\;\; R_c
\label{skbiq}
\end{eqnarray}
where $R_c$ is termed as communication range. 

\item
the signal from any unintended transmitter $t_{i,k}$ is received at
$r_{i,j}$ with an SNR less than a certain threshold $\gamma_i$, termed
as the interference threshold. This translates to
\begin{multline}
D(t_{i,k},r_{i,j}) 
\geqslant \left(\frac{P}{N_0 \gamma_i}\right)^\frac{1}{\alpha}
 =: R_i \\ \;\;\forall\;\; k=1,\ldots,M_i, \;k \neq j 
\label{djspd}
\end{multline}
where $R_i$ is termed as interference range. Note that $0 < \gamma_i <
\gamma_c$, thus $R_i > R_c$.

\end{enumerate}
The physical model of our system is denoted by $\Phi(N,({\mathbf
r}_1,\ldots,{\mathbf r}_N),P, \gamma_c,\gamma_i,\alpha,N_0)$.  

A schedule $\Psi(\cdot)$ is {\it feasible} if it satisfies the
following:
\begin{enumerate}
\item
Operational constraint: A node must not perform multiple operations in
a single time slot.
\begin{multline}
\{t_{i,j},r_{i,j}\} \cap \{t_{i,k},r_{i,k}\} = \phi \;\;\forall\;\;
i=1,\ldots,C \\\;\;\forall\;\; 1 \leqslant j < k \leqslant M_i
\label{xvnid}
\end{multline}

\item
Communication range constraint: Every receiver is within the
communication range of its intended transmitter.
\begin{multline}
D(t_{i,j},r_{i,j}) \leqslant R_c \;\forall i=1,\ldots,C \;\forall
 j=1,\ldots,M_i
\label{jdvot}
\end{multline}

\end{enumerate}

A schedule $\Psi(\cdot)$ is {\it exhaustive} if it satisfies the
following:
\begin{multline}
D(j,k) \leqslant R_c \Rightarrow j \rightarrow k \in \bigcup_{i=1}^C {\mathcal S}_i
 \;\;\mbox{and}\;\; k \rightarrow j \in \bigcup_{i=1}^C {\mathcal S}_i
 \\\;\; \forall \;\; 1 \leqslant j < k \leqslant N
\label{vuufl}
\end{multline}

A schedule $\Psi(\cdot)$ is {\it conflict-free} in terms of SINR, if
the SINR at every intended receiver does not drop below the
communication threshold.
\begin{eqnarray}
\mbox{SINR}_{r_{i,j}} \geqslant \gamma_c \;\;\forall\;\; i=1,\ldots,C, 
\;\;\forall\;\; j=1,\ldots,M_i
\label{pozxm}
\end{eqnarray}

\subsection{Graph-Based Scheduling}

The traditional approach in designing reuse schedules is to use a
graph model of the network and study the set of edges 
\cite{behzad_rubin__performance_graph}
\cite{gronkvist_hansson__comparison_between}.  The STDMA
network $\Phi(\cdot)$ is modeled by a directed graph ${\mathcal
G}({\mathcal V},{\mathcal E})$, where $\mathcal V$ is the set of
vertices and $\mathcal E$ is the set of edges. Let $\mathcal V =
\{v_1,v_2,\ldots,v_N\}$, where vertex $v_j$ represents the $j^{th}$
node in $\Phi(\cdot)$. In general, ${\mathcal E} = {\mathcal E}_c
\,\dotcup\, {\mathcal E}_i$, where ${\mathcal E}_c$ and ${\mathcal
E}_i$ denote the set of communication and interference edges
respectively. If node $k$ is within node $j$'s communication range,
then there is a communication edge from $v_j$ to $v_k$, denoted by
$v_j \stackrel{c}{\rightarrow} v_k$. If node $k$ is outside node $j$'s
communication range but within its interference range, then there is
an interference edge from $v_j$ to $v_k$, denoted by $v_j
\stackrel{i}{\rightarrow} v_k$. Thus, the mapping from $\Phi(\cdot)$
to $\mathcal G(\cdot)$ can be described as follows:

\begin{eqnarray*}
D(j,k) \leqslant R_c &\Rightarrow& v_j \stackrel{c}{\rightarrow} v_k \in 
 {\mathcal E}_c \;\;\mbox{and}\;\; v_k \stackrel{c}{\rightarrow} v_j \in 
 {\mathcal E}_c\\
R_c < D(j,k) \leqslant R_i &\Rightarrow& v_j \stackrel{i}{\rightarrow}  v_k \in 
 {\mathcal E}_i \;\;\mbox{and}\;\; v_k \stackrel{i}{\rightarrow}  v_j \in 
 {\mathcal E}_i
\end{eqnarray*}

A communication or an interference edge from $v_j$ to $v_k$ will be
denoted by $v_j \rightarrow v_k$.  The subgraph ${\mathcal
G}_c({\mathcal V},{\mathcal E}_c)$ consisting of communication edges
only is termed as the {\it communication graph}.

The schedule $\Psi(\cdot)$ is then designed from the graph $\mathcal
G(\cdot)$. Specifically, an STDMA link scheduling algorithm is
equivalent to assigning a unique color to every communication edge in
the graph, such that source-destination pairs corresponding to
communication edges with the same color transmit simultaneously in a
particular time slot. The traditional method for link assignment
requires that two communication edges $v_i \stackrel{c}{\rightarrow}
v_j$ and $v_k \stackrel{c}{\rightarrow} v_l$ can be colored the same
iff:
\begin{enumerate}
\renewcommand{\theenumi}{\roman{enumi}}
\item
vertices $v_i$, $v_j$, $v_k$, $v_l$ are all mutually distinct, i.e.,
there is no {\it primary edge conflict,} and
\label{lgcnn}

\item
$v_i \rightarrow v_l \not\in \mathcal G(\cdot)$ and $v_k \rightarrow v_j
\not\in \mathcal G(\cdot)$, i.e, there is no {\it secondary edge conflict}.
\label{plbbz}
\end{enumerate}
The first criterion is based on the operational constraint. The second
criterion states that a node cannot receive a packet while neighboring
nodes are transmitting.

Graph-Based scheduling algorithms utilize various graph coloring
methodologies to obtain a non-conflicting schedule, i.e., a schedule
devoid of primary and secondary edge conflicts.  To maximize the
throughput of an STDMA network, graph-based scheduling algorithms seek
to minimize the total number of colors used to color all the
communication edges of $\mathcal G(\cdot)$.

\subsection{Limitations of Graph-Based Scheduling Algorithms}
\label{bcprd}

Observe that Criteria \ref{lgcnn}) and \ref{plbbz}) are not sufficient
to guarantee that the resulting schedule $\Psi(\cdot)$ is
conflict-free. The link assignments that fulfill the above criteria do
not necessarily satisfy the SINR condition (\ref{pozxm}).

Importantly, graph-based scheduling algorithms do not maximize the
throughput of an STDMA network because:
\begin{enumerate}

\item
\label{mxfiw}
Due to hard-thresholding based on communication and interference
radii, graph-based scheduling algorithms can lead to high cumulative
interference at a receiver \cite{behzad_rubin__performance_graph}
\cite{gronkvist_hansson__comparison_between}.  This is because the
SINR at receiver $r_{i,j}$ decreases with an increase in the number of
concurrent transmissions $M_i$, while $R_c$ and $R_i$ have been
defined for a single transmission only.
\begin{figure}[thbp]
\centering
\includegraphics[width=3.4in]{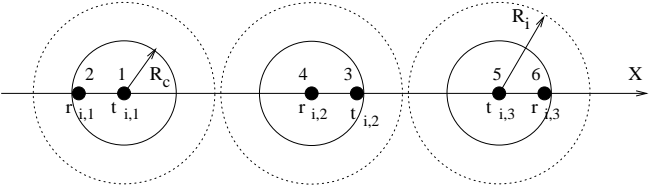}
\caption{Graph-Based algorithms can lead to high cumulative interference.}
\label{nvxye}
\end{figure}
For example, consider Figure \ref{nvxye} with six labeled nodes whose
coordinates are $1 \equiv (-360,0)$, $2 \equiv (-450,0)$, $3 \equiv
(90,0)$, $4 \equiv (0,0)$, $5 \equiv (360,0)$ and $6 \equiv (450,0)$.
The system parameters are $P=10$ mW, $\alpha = 4$, $N_0=-90$ dBm,
$\gamma_c = 20$ dB and $\gamma_i=10$ dB, which yields $R_c=100$ m and
$R_i=177.8$ m.  A graph-based scheduling algorithm will typically
schedule the transmissions $1 \rightarrow 2$, $3 \rightarrow 4$ and $5
\rightarrow 6$ in the same time slot, say the $i^{th}$ time slot, since
the resulting graph coloring is devoid of primary and secondary edge
conflicts. However, our computations show that the SINRs at receivers
$r_{i,1}$, $r_{i,2}$ and $r_{i,3}$ are $21.26$ dB, $18.42$ dB and
$19.74$ dB respectively.  From the physical interference model,
transmission $t_{i,1} \rightarrow r_{i,1}$ is successful, while
transmissions $t_{i,2} \rightarrow r_{i,2}$ and $t_{i,3} \rightarrow
r_{i,3}$ are unsuccessful. This leads to low throughput.

\item
On the other hand, graph-based scheduling algorithms can be extremely
conservative and result in a higher number of colors
\cite{moscibroda_wattenhofer_weber__protocol_design}.
\begin{figure}[thbp]
\centering
\includegraphics[width=2.4in]{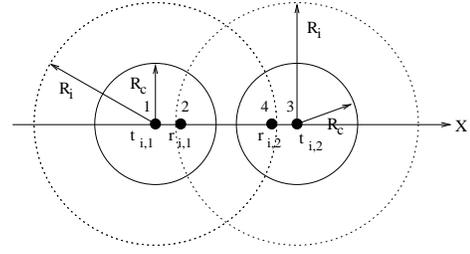}
\caption{Graph-Based algorithms can lead to higher number of colors.}
\label{hwerd}
\end{figure}
For example, with the same system parameters as in \ref{mxfiw}),
consider Figure \ref{hwerd} with four labeled nodes whose coordinates
are $1 \equiv (0,0)$, $2 \equiv (50,0)$, $3 \equiv (220,0)$ and $4
\equiv (170,0)$. Assume there are two transmission requests: $1
\rightarrow 2$ and $3 \rightarrow 4$.  If both the transmissions are
scheduled in the same slot, say the $i^{th}$ time slot, our
computations show that the SINRs at receivers $r_{i,1}$ and $r_{i,2}$
are both equal to $20.91$ dB. From the physical interference model,
both transmissions $t_{i,1} \rightarrow r_{i,1}$ and $t_{i,2}
\rightarrow r_{i,2}$ are successful, since signals levels are so high
at the receivers that strong interferences can be tolerated. However,
due to secondary edge conflicts, a graph-based scheduling algorithm
will schedule the above transmissions in different slots, thus
decreasing the throughput.

\item
Graph-based scheduling algorithms are not geography-aware, i.e., they
determine a schedule without being cognizant of the exact positions of
the transmitters and receivers.

\end{enumerate}

\subsection{Problem Formulation}
\label{pthme}

In STDMA, we construct a graph model ${\mathcal G}({\mathcal
V},{\mathcal E}_c \,\dotcup\, {\mathcal E}_i)$ of the physical network
$\Phi(\cdot)$.  The communication graph ${\mathcal G}_c({\mathcal
V},{\mathcal E}_c)$ is an approximation of $\Phi(\cdot)$, while the
two-tier graph ${\mathcal G}({\mathcal V},{\mathcal E}_c \,\dotcup\,
{\mathcal E}_i)$ is a better approximation of $\Phi(\cdot)$.  From
$\Phi(\cdot)$ and ${\mathcal G}_c(\cdot)$, one can exhaustively
determine the STDMA schedule which yields the highest throughput
according to the physical interference model. However, this is a
combinatorial optimization problem of prohibitive complexity
$(O(|{\mathcal E}_c|^{|{\mathcal E}_c|}))$ and is thus computationally
infeasible.

To overcome these problems, we propose a new suboptimal algorithm for
STDMA link scheduling based on a more realistic physical interference
model.  Our algorithm is based on the communication graph model
${\mathcal G}_c({\mathcal V},{\mathcal E}_c)$ as well as SINR
computations.

To evaluate the performance of our algorithm and compare it with
existing suboptimal STDMA link scheduling algorithms, we define the
following metric: spatial reuse.  Consider the STDMA link schedule
$\Psi(\cdot)$ for the network $\Phi(\cdot)$. Under the physical
interference model, the transmission $t_{i,j} \rightarrow r_{i,j}$ is
successful iff (\ref{mnmnx}) is satisfied.  The {\it spatial reuse} of
the schedule $\Psi(\cdot)$ is defined as the average number of
successfully received packets per time slot in the STDMA
schedule. Thus
\begin{multline}
\mbox{Spatial Reuse $\;=\; \sigma$} =
 \frac{\sum_{i=1}^C\sum_{j=1}^{M_i}I({\mbox{SINR}}_{r_{i,j}}\geqslant\gamma_c)}{C} 
\label{qpjkk}
\end{multline}
where $I(A)$ denote the indicator function for event $A$, i.e.,
$I(A)=1$ if event $A$ occurs, $I(A)=0$ if event $A$ does not occur.

The essence of STDMA is to have a reasonably large number of
concurrent and successful transmissions.  For an STDMA network which
is operational for a long period of time, say $L$ slots, the total
number of successfully received packets is $L\sigma$. Thus, a high
value of spatial reuse\footnote{Note that spatial reuse in our
  system model is analogous to spectral efficiency in digital
  communication systems.} directly translates to higher long-term total
network throughput and the number of colors $C$ is relatively
unimportant.  Hence, spatial reuse turns out to be a crucial metric
for the comparison of different STDMA algorithms.

We seek low complexity conflict-free STDMA link scheduling algorithms
with high spatial reuse. We only consider STDMA schedules which are
feasible and exhaustive. Thus, our schedules satisfy (\ref{xvnid}),
(\ref{jdvot}), (\ref{vuufl}) and (\ref{pozxm}).

\subsection{Comparison with SINR-Based Scheduling Algorithms}

In cognizance of our system model and performance metric, our work is
reasonably different from existing SINR-based STDMA link scheduling
algorithms.

The works in \cite{brar_blough_santi__computationally_efficient}
\cite{moscibroda_wattenhofer__complexity_connectivity} focus on
minimizing the schedule length, which does not necessarily translate
to high network throughput. On the other hand, spatial reuse directly
corresponds to network throughput capacity
\cite{gupta_kumar__capacity_wireless}.  Power-controlled algorithms
can lead to excessively high transmit power (for example, Line 16 in
Algorithm 1 \cite{moscibroda_wattenhofer__complexity_connectivity}),
which is impractical since all wireless routers have constraints on
maximum transmit power. So, similar to
\cite{kodialam_nandagopal__characterizing_achievable}, we consider
uniform transmit power at all wireless routers.  Existing works on
SINR-based link scheduling, which are usually in the context of 802.11
wireless networks \cite{brar_blough_santi__computationally_efficient}
\cite{kim_lim_hou__improving_spatial} and wireless mesh networks
\cite{lim_lim_hou__coordinate_based}
\cite{kodialam_nandagopal__characterizing_achievable}, consider many
practical aspects of the underlying communication protocol and network
architecture. Consequently, their system models are quite different
from our pure STDMA network model.

To the best of our knowledge, this is the first attempt to develop a
centralized algorithm for SINR-based link scheduling in a pure STDMA
wireless ad hoc network with uniform power assignment.  Also, this
work is different from previous works due to the focus on spatial
reuse.

Hence, we compare the performance of our algorithm with existing
graph-based algorithms only.

\section{SINR-Based Link Scheduling Algorithm}
\label{lxvym}

\subsection{Structure}

We first describe the essential features of STDMA link scheduling
algorithms. The core of every link scheduling algorithm consists
of the following modules:
\begin{enumerate}

\item
An order in which communication edges are considered for coloring.

\item
A function which determines the set of all existing colors which can
be assigned to the edge under consideration without violating the
problem constraints.

\item
A {\it BestColor} rule to determine which conflict-free color to
assign to the edge under consideration.
\end{enumerate}

The second module considers only operational and communication range
constraints in graph-based scheduling algorithms. However, in
the SINR-based link scheduling algorithm that we propose, SINR
constraints are also taken into account. Note that this function is
completely described by the problem definition and does not vary from
algorithm to algorithm. The ordering of edges for coloring and the
BestColor rule play a significant role in determining the performance
and computational complexity of an STDMA scheduling algorithm.

\subsection{Motivation}

Recall that graph-based models are inadequate to design efficient link
schedules under the physical interference model and brute-force
computation of an optimal link schedule that maximizes spatial reuse
is prohibitively complex (see Sections \ref{bcprd} and \ref{pthme}).
Motivated by techniques from matroid theory
\cite{lawler__combinatorial_optimization}, we develop a
computationally feasible algorithm with demonstrably high spatial
reuse.  The essence of our algorithm is to partition the set of
communication edges into subsets (forests) and color the edges in each
subset sequentially. The edges in each forest are considered in a
random order for coloring, since randomized algorithms are known to
outperform deterministic algorithms, esp. when the characteristics of
the input are not known apriori
\cite{motwani_raghavan__randomized_algorithms}.

A similar matroid-based network partitioning technique is used in
\cite{brzezinski_zussman_modiano__enabling_distributed} to generate
high capacity subnetworks for a distributed throughput maximization
problem in wireless mesh networks.  Techniques from matroid theory
have also been employed to develop efficient heuristics for NP-hard
combinatorial optimization problems in fields such as distributed
computer systems \cite{ramalingom_thulasiraman_das__matroid_theoretic}
and linear network theory \cite{petersen__investigating_solvability}.

\subsection{ConflictFreeLinkSchedule Algorithm}
\label{xbpyf}

Our proposed SINR-based link scheduling algorithm is
ConflictFreeLinkSchedule, which considers the communication graph
${\mathcal G}_c({\mathcal V},{\mathcal E}_c)$ and is described in
Algorithm \ref{hidwo}.

In Phase 1 (Line 3), we label all the vertices randomly. Specifically,
if ${\mathcal G}_c(\cdot)$ has $v$ vertices, we perform a random
permutation of the sequence $(1,2,\ldots,v)$ and assign these labels
to vertices with indices $1,2,\ldots,v$ respectively.

In Phase 2 (Line 4), the communication graph ${\mathcal G}_c(\cdot)$
is decomposed into what are called as out-oriented and in-oriented
graphs $T_1,T_2,\ldots,T_k$
\cite{ramanathan_lloyd__scheduling_algorithms}. Each $T_i$ is a forest
and every edge of ${\mathcal G}_c(\cdot)$ is in exactly one of the
$T_i$'s. This decomposition is achieved by partitioning graph
$G_c(\cdot)$, the undirected equivalent of ${\mathcal G}_c(\cdot)$,
into undirected forests. The number of forests can be minimized by
using techniques from Matroid theory ($k$-forest problem,
\cite{gabow_westermann__forests_frames}). However, this optimal
decomposition requires extensive computation. Hence, we adopt the
speedier albeit non-optimal approach of using successive breadth first
searches to decompose $G_c(\cdot)$ into undirected forests. Each
undirected forest is further mapped to two directed forests. In one
forest, the edges in every connected component point away from the
root and every vertex has at most one incoming edge, thus producing an
out-oriented graph. In the other forest, the edges in every connected
component point toward the root and every vertex has at most one
outgoing edge, thus producing an in-oriented graph\footnote{An
  in-oriented graph is also constructed by Algorithm 1 in
  \cite{moscibroda_wattenhofer__complexity_connectivity} to determine
  a link schedule in a power-controlled STDMA network.}.

In Phase 3 (Lines 5-14), the oriented graphs are considered
sequentially.  For each oriented graph, vertices are considered in
increasing order by label\footnote{In essence, the edges are scanned
  in a random order, since labeling is random.} and the unique edge
associated with each vertex is colored using the
FirstConflictFreeColor function.

\begin{algorithm}
\caption{ConflictFreeLinkSchedule}
\label{hidwo}
\begin{algorithmic}[1]
\STATE {\bf input:} Physical network
$\Phi(\cdot)$, communication graph ${\mathcal G}_c(\cdot)$
\STATE {\bf output:} A coloring $C: {\mathcal E}_c \rightarrow \{1,2,\ldots\}$
\STATE label the vertices of ${\mathcal G}_c$ randomly
\STATE use successive breadth first searches to partition ${\mathcal G}_c$ into oriented graphs
$T_i$, $1 \leqslant i \leqslant k$
\FOR{$i \leftarrow 1 \mbox{ to } k$}
\FOR{$j \leftarrow 1 \mbox{ to } n$}
\IF{$T_i$ is out-oriented}
\STATE let $x=(s,d)$ be such that $L(d)=j$
\ELSE
\STATE let $x=(s,d)$ be such that $L(s)=j$
\ENDIF
\STATE $C(x) \leftarrow \mbox{FirstConflictFreeColor}(x)$
\ENDFOR
\ENDFOR
\end{algorithmic}
\end{algorithm}

The FirstConflictFreeColor function is explained in Algorithm
\ref{pecdi}. For the edge under consideration $x$, it discards any
color that has an edge with a primary conflict with $x$. Now, we
consider the residual set of conflict-free colors sequentially. We
choose the first conflict-free color such that the resulting SINRs at
the receiver of $x$ and the receivers of all co-colored edges exceed
the communication threshold $\gamma_c$. If no such color is found, we
assign a new color to $x$. Thus, this function guarantees that the
ensuing schedule is conflict-free in terms of SINR. Since we choose
the first SINR-compliant color and not the `best' SINR-compliant color
according to some BestColor rule, the computational complexity of this
function is lower than that of any other function which checks the
SINRs at the receivers of all co-colored edges.

\begin{algorithm}
\caption{integer FirstConflictFreeColor($x$)}
\label{pecdi}
\begin{algorithmic}[1]
\STATE {\bf input:} Physical network $\Phi(\cdot)$, communication graph 
${\mathcal G}_c(\cdot)$
\STATE {\bf output:} A conflict-free color
\STATE ${\mathcal C} \leftarrow \mbox{set of existing colors}$
\STATE ${\mathcal C}_c \leftarrow \{C(h):h \in {\mathcal E}_c$, $h$ is colored, $x$ and $h$ have a primary edge conflict$\}$
\STATE ${\mathcal C}_{cf} = {\mathcal C} \setminus {\mathcal C}_c$
\FOR{$i \leftarrow 1 \mbox{ to } |{\mathcal C}_{cf}|$}
\STATE $r \leftarrow i^{th} \mbox{ color in } {\mathcal C}_{cf}$
\STATE $E_r \leftarrow \{h:h \in {\mathcal E}_c, C(h)=r\}$
\STATE $C(x) \leftarrow r$
\IF{SINR at all receivers of $E_r \cup \{x\}$ exceed $\gamma_c$}
\STATE return $r$
\ENDIF
\ENDFOR
\STATE return $|\mathcal C|+1$
\end{algorithmic}
\end{algorithm}

\section{Performance Results}
\label{cymoe}

\subsection{Simulation Model}
\label{zxptb}

In our simulation experiments, the location of every node is generated
randomly, using a uniform distribution for its $X$ and $Y$
coordinates, in the deployment area. For a fair comparison of our
algorithm with the Truncated Graph-Based Scheduling Algorithm
\cite{behzad_rubin__performance_graph}, we assume that the deployment
region is a circular region of radius $R$. Thus, if $(X_j,Y_j)$ are
the Cartesian coordinates of the $j^{th}$ node, $j=1,\ldots,N$, then
$X_j \sim U[-R,R]$ and $Y_j \sim U[-R,R]$ subject to $X_j^2 + Y_j^2
\leqslant R^2$. Equivalently, if $(R_j,\Theta_j)$ are the polar
coordinates of the $j^{th}$ node, then $R_j^2 \sim U[0,R^2]$ and
$\Theta_j \sim U[0,2\pi]$.  After generating random positions for $N$
nodes, we have complete information of $\Phi(\cdot)$. Using
(\ref{skbiq}) and (\ref{djspd}), we compute the communication and
interference radii, and then map the network $\Phi(\cdot)$ to the
two-tier graph ${\mathcal G}({\mathcal V},{\mathcal E}_c \,\dotcup\,
{\mathcal E}_i)$.  Once the link schedule is computed by every
algorithm, the spatial reuse is computed using (\ref{qpjkk}).  We use
two sets of values for system parameters $P$, $\gamma_c$, $\gamma_i$,
$\alpha$, $N_0$, $N$ and $R$, which are prototypical values of system
parameters in wireless networks \cite{kim_lim_hou__improving_spatial}
and describe them in Section \ref{mclgs}.  For a given set of system
parameters, we calculate the spatial reuse by averaging this quantity
over one thousand randomly generated networks.  Keeping all other
parameters fixed, we observe the effect of increasing the number of
nodes $N$ on the spatial reuse $\sigma$.

In our experiments, we compare the performance of the following algorithms:
\begin{enumerate}
\item
ArboricalLinkSchedule \cite{ramanathan_lloyd__scheduling_algorithms}
(ALS)

\item 
  Truncated Graph-Based Scheduling Algorithm\footnote{In Truncated
    Graph-Based Scheduling Algorithm, for the computation of optimal
    number of transmissions $M^*$, we follow the exact method
    described in \cite{behzad_rubin__performance_graph}. Since $0 <
    \xi < \frac{N_0}{P}$, we assume that $\xi = 0.9999\frac{N_0}{P}$
    and compute successive Edmundson-Madansky (EM) upper bounds
    \cite{madansky__inequalities_stochastic}
    \cite{dokov_morton__higher_order} till the difference between
    successive EM bounds is less than $0.3\%$. We have experimentally
    verified that only high values of $\xi$ lead to reasonable values
    for $M^*$, whereas low values of $\xi$, say $\xi =
    0.1\frac{N_0}{P}$, lead to the extremely conservative value of
    $M^*=1$ in most cases.}  \cite{behzad_rubin__performance_graph}
  (TGSA)

\item
ConflictFreeLinkSchedule (CFLS)

\end{enumerate}

\subsection{Performance Comparison}
\label{mclgs}

\begin{figure}[thbp]
\centering
\includegraphics[width=3.6in]{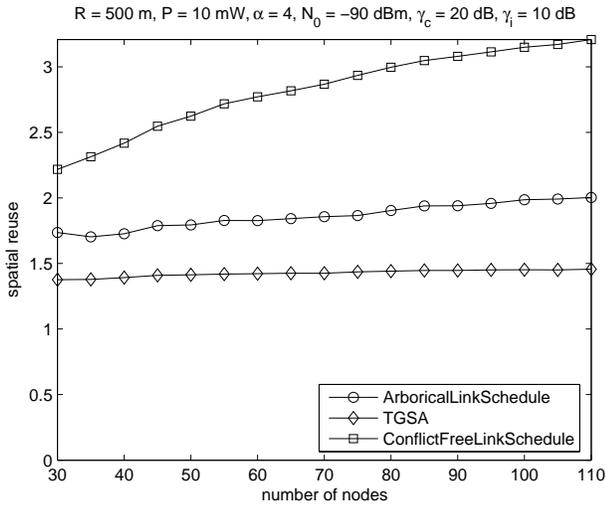}
\caption{Spatial reuse vs. number of nodes for Experiment 1.}
\label{zowyx}
\end{figure}

In our first set of experiments (Experiment 1), we assume that $R=500$
m, $P=10$ mW, $\alpha = 4$, $N_0=-90$ dBm, $\gamma_c = 20$ dB and
$\gamma_i=10$ dB. Thus, $R_c=100$ m and $R_i=177.8$ m.  We vary the
number of nodes from 30 to 110 in steps of 5. Figure \ref{zowyx} plots
the spatial reuse vs. number of nodes for all the algorithms.

\begin{figure}[thbp]
\centering
\includegraphics[width=3.6in]{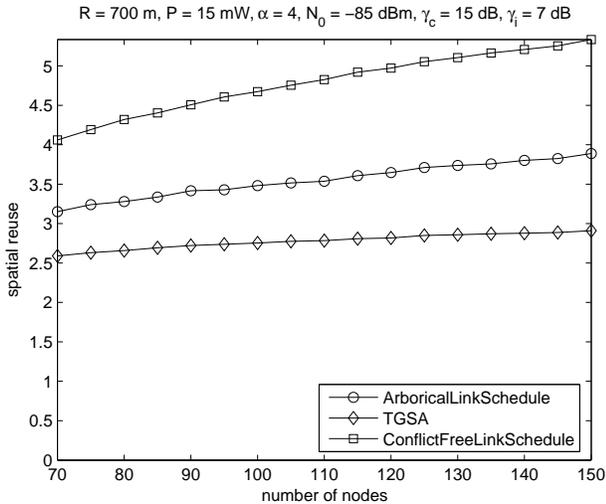}
\caption{Spatial reuse vs. number of nodes for Experiment 2.}
\label{uusvr}
\end{figure}

In our second set of experiments (Experiment 2), we assume that
$R=700$ m, $P=15$ mW, $\alpha = 4$, $N_0=-85$ dBm, $\gamma_c = 15$ dB
and $\gamma_i=7$ dB. Thus, $R_c=110.7$ m and $R_i=175.4$ m.  We vary
the number of nodes from 70 to 150 in steps of 5. Figure \ref{uusvr}
plots the spatial reuse vs. number of nodes for all the algorithms.

For the ALS algorithm, we observe that spatial reuse increases very
slowly with increasing number of nodes.

For the TGSA algorithm, we observe that spatial reuse is $18$-$27\%$
lower than that of ALS. A plausible explanation for this behavior is
as follows.  The basis for TGSA is the computation of $M^*$, the
optimal number of transmissions in every slot.  $M^*$ is determined by
maximizing a lower bound on the expected number of successful
transmissions in a time slot.  Since the partitioning of a maximal
independent set of communication arcs into subsets of cardinality at
most $M^*$ is arbitrary and not geography-based, there could be
scenarios where the transmissions scheduled in a subset are in the
vicinity of each other, resulting in moderate to high interference.
In essence, maximizing this lower bound does not necessarily translate
to maximizing the number of successful transmissions in a time slot.
Also, due to its design, the TGSA algorithm yields higher number of
colors compared to ALS.

For our CFLS algorithm, we observe that spatial reuse increases
steadily with increasing number of nodes and is consistently
$25$-$50\%$ higher than the spatial reuse of ALS and TGSA.

\subsection{Performance Comparison under Realistic Channel
Conditions}

In a realistic wireless environment, channel impairments like
multipath fading and shadowing affect the received SINR at a receiver
\cite{sklar__rayleigh_fading}.  In this section, we compare the
performance of the ALS, TGSA and CFLS algorithms in a wireless channel
which experiences Rayleigh fading and lognormal shadowing.

In the absence of fading and shadowing, the SINR at receiver $r_{i,j}$
is given by (\ref{xvmyw}). We assume that every algorithm (ALS, TGSA
and CFLS) considers only path loss in the channel prior to
constructing the two-tier graph ${\mathcal G}({\mathcal V},{\mathcal
  E}_c \,\dotcup\, {\mathcal E}_i)$ and computing the link schedule.
However, when evaluating the performance of each algorithm, we take
into account the fading and shadowing channel gains between every pair
of nodes. Specifically, for computing the spatial reuse using
(\ref{qpjkk}), we assume that the (actual) SINR at receiver $r_{i,j}$
is given by
\begin{multline*}
{\mbox{SINR}}_{r_{i,j}} = 
 \frac{\frac{P}{D^{\alpha}(t_{i,j},r_{i,j})}V(t_{i,j},r_{i,j})10^{W(t_{i,j},r_{i,j})}}{N_0+\sum_{\stackrel{k=1}{k\neq j}}^{M_i}\frac{P}{D^{\alpha}(t_{i,k},r_{i,j})}V(t_{i,k},r_{i,j})10^{W(t_{i,k},r_{i,j})}}
\end{multline*}
where random variables $V(\cdot)$ and $W(\cdot)$ correspond to channel
gains due to Rayleigh fading and lognormal shadowing respectively. We
assume that $\{V(k,l)|1\leqslant k,l \leqslant N, k \neq l\}$ are
independent and identically distributed (i.i.d.) random variables with
probability density function\footnote{$u(\cdot)$ is the unit step
  function.} (pdf)
\cite{tse_viswanath__fundamentals_wireless}
\begin{eqnarray*}  
f_V(v) &=& \frac{1}{\sigma_V^2}e^{\frac{-v}{\sigma_V^2}}u(v)
\end{eqnarray*}
and $\{W(k,l)|1\leqslant k,l \leqslant N, k\neq l\}$ are i.i.d. zero mean
Gaussian random variables with pdf
\cite{goldsmith__wireless_communications}
\begin{eqnarray*}
f_W(w) &=& \frac{1}{\sqrt{2\pi}\sigma_W}e^{\frac{-w^2}{2\sigma_W^2}}
\end{eqnarray*}
Random variables $V(\cdot)$ and $W(\cdot)$ are independent of each
other and also independent of the node locations.

\begin{figure}[thbp]
\centering
\includegraphics[width=3.6in]{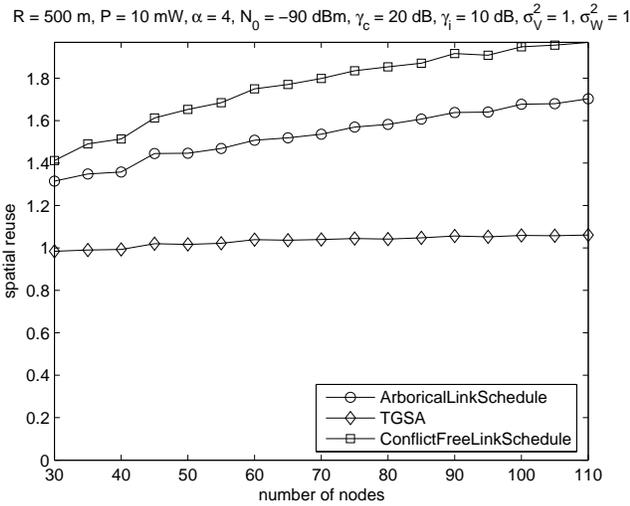}
\caption{Spatial reuse vs. number of nodes for Experiment 1 under 
multipath fading and shadowing channel conditions.}
\label{cvcvb}
\end{figure}

\begin{figure}[thbp]
\centering
\includegraphics[width=3.6in]{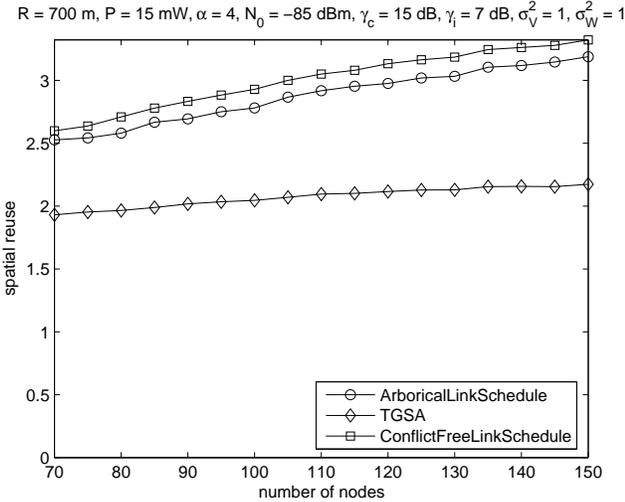}
\caption{Spatial reuse vs. number of nodes for Experiment 2 under 
multipath fading and shadowing channel conditions.}
\label{vcvcb}
\end{figure}

Our simulation model and experiments are exactly as described in
Sections \ref{zxptb} and \ref{mclgs}. In our simulations, we assume
$\sigma_V^2 = \sigma_W^2 = 1$.  For Experiment 1, Figure \ref{cvcvb}
plots the spatial reuse vs. number of nodes for all the algorithms.
For Experiment 2, Figure \ref{vcvcb} plots spatial reuse vs. number of
nodes for all the algorithms.

From Figures \ref{zowyx}, \ref{uusvr}, \ref{cvcvb} and \ref{vcvcb}, we
observe that spatial reuse decreases by $20$-$40\%$ in a channel
experiencing multipath fading and shadowing effects. A plausible
explanation for this observation is as follows. Since the channel
gains between every pair of nodes are independent of each other, it is
reasonable to assume that the interference power at a typical receiver
remains almost the same as in the non-fading case. This is because,
even if the power received from few unintended transmitters is low,
the power received from other unintended transmitters will be high (on
an average); thus the interference power remains constant.
Consequently, the change in SINR is determined by the change in
received signal power only. If the received signal power is higher
compared to the non-fading case, the transmission is anyway successful
and spatial reuse remains unchanged (see (\ref{qpjkk})). However, if
the received signal power is lower, the transmission is now
unsuccessful and spatial reuse decreases. Hence, on an average, the
spatial reuse decreases.

Finally, from Figures \ref{cvcvb} and \ref{vcvcb}, we observe that our CFLS
algorithm achieves $5$-$17\%$ higher spatial reuse than the ALS
algorithm and $40$-$80\%$ higher spatial reuse than the TGSA
algorithm, under realistic channel conditions.

\section{Analytical Results}
\label{mbxgt}

In this section, we derive upper bounds on the running time complexity
(computational complexity) of the ConflictFreeLinkSchedule algorithm.
We will use the following notation with respect to the communication
graph ${\mathcal G}_c({\mathcal V},{\mathcal E}_c)$:
\begin{eqnarray*}
e &=& \mbox{number of communication edges}\\
v &=& \mbox{number of vertices}\\
\theta &=& \mbox{thickness of the graph}\\
&:=& \mbox{minimum number of graphs into which the}\\
&& \mbox{undirected equivalent of ${\mathcal G}_c(\cdot)$ can be partitioned}
\end{eqnarray*}

\begin{figure}
\centering
\subfigure[Experiment 1]
{
\label{ywhlv}
\includegraphics[width=3.35in]{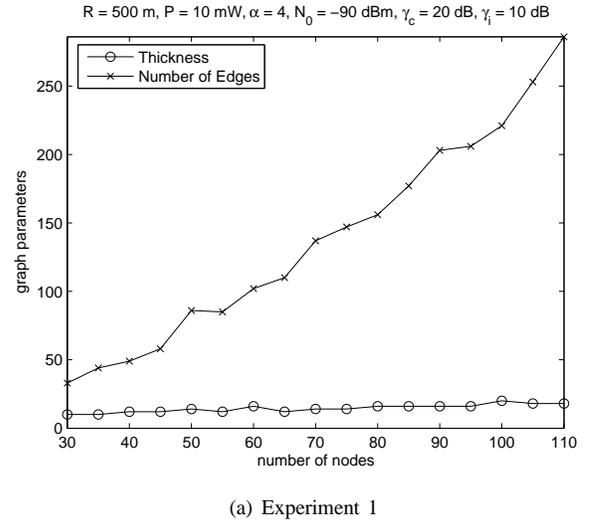}
}
\hspace{0.2in}
\subfigure[Experiment 2]
{
\label{drjnb}
\includegraphics[width=3.35in]{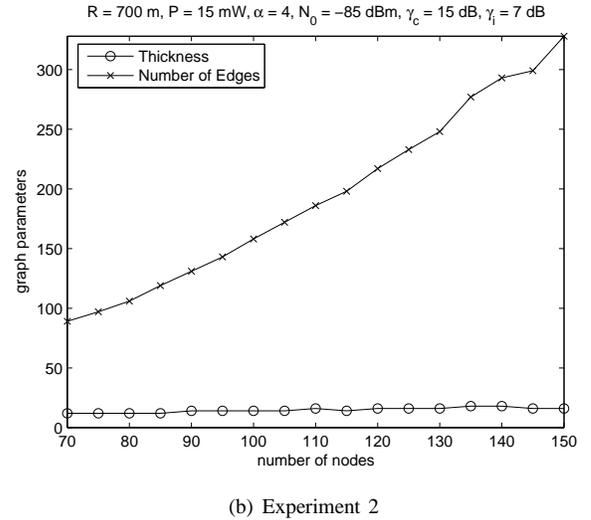}
}
\caption{Comparison of thickness and number of edges with number of vertices.}
\label{wvlpt}
\end{figure}

Before we prove our results, it is instructive to observe Figure
\ref{wvlpt}, which shows the variation of $\theta$ and $e$ with $v$
for the two experiments described in Section \ref{mclgs}.  Since
determining the thickness of a graph is NP-hard
\cite{mutzel_odenthal__thickness_graphs}, each value of $\theta$ in
Figure \ref{wvlpt} is an upper bound on the actual thickness based on
the number of forests into which the undirected equivalent of the
communication graph has been decomposed using successive breadth first
searches.  We observe that the graph thickness increases very slowly
with the number of vertices, while the number of edges increases
super-linearly with the number of vertices.

\begin{lemma}
\label{apmxr}
An oriented graph $T$ can be colored using no more than $O(v)$ colors
using ConflictFreeLinkSchedule.
\end{lemma}
\begin{proof}
Since an oriented graph with $v$ vertices has at most $v$ edges, the
edges of $T$ can be colored with at most $v$ colors.
\end{proof}

\vspace{1.5ex}

\begin{remark}
The number of colors obtained by our algorithm depends not only on the
graph representation of the STDMA network, but also on the positions
of the nodes and the values of $P$, $\gamma_c$, $\gamma_i$, $\alpha$
and $N_0$. Since our algorithm is both graph-based and SINR-based
(hybrid), it is not possible to obtain a tighter upper bound in Lemma
\ref{apmxr}.
\end{remark}

\vspace{1.5ex}

\begin{lemma}
For an oriented graph $T$, the running time of
ConflictFreeLinkSchedule is $O(v^2)$.
\label{dolfb}
\end{lemma}
\begin{proof}
Assuming that an element can be chosen randomly and uniformly from a
finite set in unit time (Chapter 1,
\cite{motwani_raghavan__randomized_algorithms}), the running time of
Phase 1 can be shown to be $O(v)$.  Since there is only one oriented
graph, Phase 2 runs in time $O(1)$. In Phase 3, the unique edge
associated with the vertex under consideration is assigned a color
using FirstConflictFreeColor. From Lemma \ref{apmxr}, the size of the
set of colors to be examined $|{\mathcal C}_c \cup {\mathcal C}_{cf}|$
is $O(v)$.  In FirstConflictFreeColor, the SINR is checked only once
for every colored edge in the set $\bigcup_{i=1}^{|{\mathcal C}_{cf}|}
E_i$ and at most $v$ times for the edge under consideration $x$. With
a careful implementation, FirstConflictFreeColor runs in time $O(v)$.
So, the running time of Phase 3 is $O(v^2)$.  Thus, the total running
time is $O(v^2)$.
\end{proof}

\vspace{3ex}

\begin{theorem}
For an arbitrary graph $\mathcal G$, the running time of
ConflictFreeLinkSchedule is $O(ev\log v + ev\theta)$.
\end{theorem}
\begin{proof}
Assuming that an element can be chosen randomly and uniformly from a
finite set in unit time
\cite{motwani_raghavan__randomized_algorithms}, the running time of
Phase 1 can be shown to be $O(v)$.  For Phase 2, the optimal
partitioning technique of \cite{gabow_westermann__forests_frames}
based on Matroids can be used to partition the communication graph
${\mathcal G}_c$ into at most $6\theta$ oriented graphs in time
$O(ev\log v)$.  Thus, $k \leqslant 6\theta$ holds for Phase 3.  From
Lemma \ref{dolfb}, it follows that the first oriented graph $T_1$ can
be colored in time $O(v^2)$. However, consider the coloring of the
$j^{th}$ oriented graph $T_j$, where $2 \leqslant j \leqslant k$.
When coloring edge $x$ from $T_j$ using FirstConflictFreeColor,
conflicts can occur not only with the colored edges of $T_j$, but also
with the edges of the previously colored oriented graphs
$T_1,T_2,\ldots,T_{j-1}$.  This fact is exemplified in Appendix
\ref{jdvow}.  Hence, the worst-case size of the set of colors to be
examined $|{\mathcal C}_c \cup {\mathcal C}_{cf}|$ is $O(e)$. Note
that in FirstConflictFreeColor, the SINR is checked only once for
every colored edge in the set $\bigcup_{i=1}^{|{\mathcal C}_{cf}|}
E_i$ and at most $e$ times for the edge under consideration $x$. With
a careful implementation, FirstConflictFreeColor runs in time $O(e)$.
Hence, any subsequent oriented graph $T_j$ can be colored in time
$O(ev)$. Thus, the running time of Phase 3 is $O(ev\theta)$.
Therefore, the overall running time of ConflictFreeLinkSchedule is
$O(ev\log v + ev\theta)$.
\end{proof}

\section{Discussion}
\label{eiovv}

In this paper, we have developed ConflictFreeLinkSchedule, an
SINR-based link scheduling algorithm for STDMA multihop wireless ad
hoc networks under the physical interference model.  The performance
of our algorithm is superior to existing link scheduling algorithms
for STDMA networks with uniform power assignment.  A practical
experimental modeling shows that, on an average, our algorithm
achieves $40\%$ higher spatial reuse than the ArboricalLinkSchedule
\cite{ramanathan_lloyd__scheduling_algorithms} and Truncated
Graph-Based Scheduling \cite{behzad_rubin__performance_graph}
algorithms.  Since schedules are constructed offline only once and
then used by the network for a long period of time, these improvements
in performance directly translate to higher long-term network
throughput.

The computational complexity of ConflictFreeLinkSchedule is comparable
to the computational complexity of ArboricalLinkSchedule and is much
lower than the computational complexity of Truncated Graph-Based
Scheduling Algorithm. Thus, in cognizance of spatial reuse as well as
computational complexity, ConflictFreeLinkSchedule is a good candidate
for efficient SINR-based STDMA link scheduling algorithms.

We have recently developed computationally efficient algorithms for
STDMA broadcast scheduling under the physical interference model. It
would be interesting to apply techniques like simulated annealing,
genetic algorithms and neural networks to compute high spatial reuse
conflict-free STDMA link schedules.

\appendices
\section{Example of Primary Edge Conflicts with Previously Colored Oriented Graphs}
\label{jdvow}

\begin{figure}[thbp]
\centering \includegraphics[width=1.35in]{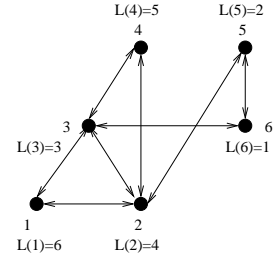}
\caption{A six-node STDMA wireless ad hoc network, its communication graph and node labels.}
\label{hxcvm}
\end{figure}

\begin{figure}[thbp]
\centering
\includegraphics[width=3.5in]{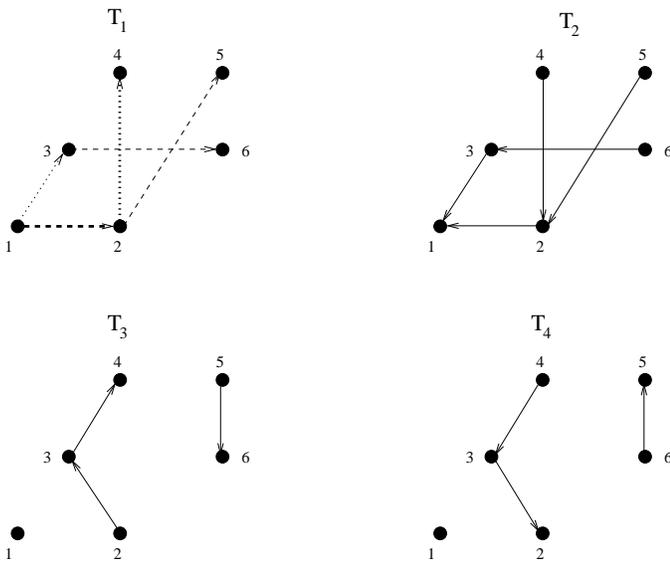}
\caption{Decomposition of Fig. \ref{hxcvm} into two out-oriented graphs $(T_1,T_3)$ and two in-oriented graphs $(T_2,T_4)$.}
\label{pndiz}
\end{figure}

Consider the six-node STDMA wireless ad hoc network shown in Figure
 \ref{hxcvm}, along with its associated communication graph ${\mathcal
 G}_c(\cdot)$ and node labels.  Using successive breadth first
 searches, ${\mathcal G}_c(\cdot)$ is partitioned into four oriented
 graphs $T_1$, $T_2$, $T_3$ and $T_4$, as shown in Figure \ref{pndiz}.
 A conflict-free coloring of the first oriented graph $T_1$ is shown
 in Table \ref{kdhfo}.  Now, when we color an edge from any other
 oriented graph, we must take into account the colors of the edges in
 $T_1$. For example:
\begin{enumerate}
\item
In $T_2$, Edge $6 \rightarrow 3$ cannot be assigned Color $2$ due to a
primary edge conflict with Edge $1 \rightarrow 3$ of $T_1$.

\item
In $T_3$, Edge $5 \rightarrow 6$ cannot be assigned Color $1$ due to a
primary edge conflict with Edge $2 \rightarrow 5$ of $T_1$.

\item
In $T_4$, Edge $3 \rightarrow 2$ cannot be assigned Color $3$ due to a
primary edge conflict with Edge $1 \rightarrow 2$ of $T_1$.

\end{enumerate}

\begin{table}[hbtp]
\begin{tabular}{|c|c|} \hline
Edge of $T_1$ & Color \\ \hline
$3 \rightarrow 6$ & 1 \\ \hline
$2 \rightarrow 5$ & 1 \\ \hline
$1 \rightarrow 3$ & 2 \\ \hline
$1 \rightarrow 2$ & 3 \\ \hline
$2 \rightarrow 4$ & 4 \\ \hline
\end{tabular}
\caption{Conflict-free coloring of first oriented graph $T_1$.}
\label{kdhfo}
\end{table}

\bibliographystyle{IEEEtran.bst}

\bibliography{s}

\end{document}